\documentstyle[sprocl,epsfig]{article}

\def\be{\begin{equation}}
\def\ee{\end{equation}}
\def\bea{\begin{eqnarray}}
\def\eea{\end{eqnarray}}

\begin{document}

\title{USING THIN FILM TARGETS FOR MUONIC ATOMS \\
 AND MUON CATALYZED FUSION STUDIES%
 \footnote{Talk given at KEK International Workshop on High
 Intensity Muon Sources, HIMUS-99.}
 }

\author{M.C. Fujiwara,$^{1,2,}$\footnote%
{E-mail:Makoto.Fujiwara@cern.ch} A. Adamczak,$^3$ J.M.~Bailey,$^4$
G.A.~Beer,$^5$ J.L.~Beveridge,$^6$ M.P.~Faifman,$^7$
T.M.~Huber,$^{8}$ P.~Kammel,$^9$ S.K.~Kim,$^{10}$
P.E.~Knowles,$^{5,11}$ A.R.~Kunselman,$^{12}$
V.E.~Markushin,$^{13}$ G.M.~Marshall,$^6$ G.R. Mason,$^5$
F.~Mulhauser,$^{6,11}$ A.~Olin,$^{5,6}$ C.~Petitjean,$^{13}$
T.A.~Porcelli,$^{5,14}$ and J.~Zmeskal$^{15}$\\ (TRIUMF Muonic
Hydrogen Collaboration)}

\address{$^1$Department of Physics, University of Tokyo, Tokyo
 113-0033 Japan,
 $^2$Department of Physics and Astronomy,
 University of British Columbia, Vancouver, BC, Canada,
 $^3$Institute of Nuclear Physics, Krakow, Poland,
 $^4$Chester Technology, Chester, England, UK,
 $^5$Department of Physics and Astronomy, University of Victoria,
 BC, Canada,
 $^6$TRIUMF, Vancouver, BC, Canada,
 $^7$Russian Research Center, Kurchatov Institute, Moscow,
 Russia,
 $^8$Department of Physics, Gustavus Adolphus College, St. Peter, MN, USA,
 $^9$Department of Physics and LBNL,
 University of California, Berkeley, CA, USA,
 $^{10}$Department of Physics, Jeonbuk National University, Jeonju
 City, S. Korea,
 $^{11}$Institute of Physics, University of Fribourg, Fribourg,
 Switzerland,
 $^{12}$Department of Physics and Astronomy,
 University of Wyoming, WY, USA,
 $^{13}$Paul Scherrer Insitute, Villigen, Switzerland,
 $^{14}$Department of Physics, University of Northern British
 Columbia, BC, Canada
 $^{15}$IMEP, Austrian Academy of
 Sciences, Vienna, Austria}


 \maketitle
 \abstracts{Studies of muonic atoms and muon catalyzed fusion have
 been conventionally done in a bulk target of gas, liquid or solid
 hydrogen isotopes.
 The use of thin film targets developed at TRIUMF
 have notable advantages in tackling some of the most important
 questions in the field, which could be further exploited at future
 high intensity muon sources. We review the technique of the thin
 film method with emphasis on recent results and a future proposal.}

\section{Introduction}

A negative muon can participate in a variety of atomic and
molecular processes. When it is introduced into a target, a muonic
atom is readily created replacing an electron, which can then form
a muonic molecule. The latter in turn can result in fusion
reactions between the nuclei if the target consists of hydrogen
isotopes, a phenomenon known as muon catalyzed fusion
($\mu$CF).\cite{review}

The use of a thin target in muon physics experiments has some
notable advantages. One general (and obvious) advantage is that it
allows, with high energy resolution, the detection of out-going
particles of the reaction, {\it e.g.} $\alpha$ particles from
fusion, or possibly electrons from $\mu - e$ conversion.
Using {\it thin film} targets, held in a vacuum without a window,
offers additional merits in muonic atom studies. When the layer
thickness is small compared to the range of the atom in that
medium, it can be extracted from the layer, producing a beam of
muonic atoms in vacuum. This allows to isolate the process of
interest from the rest of the reaction chains, offering unique
advantages over conventional targets in which usually complex and
interconnected of processes take place. There exist some
disadvantages of thin film methods which include limited
statistics due to the small fraction of muon stopping in the thin
films, and increased background from the large fraction muon
stopping in the target support and the chamber.

New high intensity muon sources, now being actively investigated
world-wide, could enhance greatly the advantages of thin targets
while alleviating their disadvantages. In this articles, we review
the thin film method,\cite{marsh92} and discuss recent results and
a future proposal in muonic atom and muon catalyzed fusion
experiments at TRIUMF.

\section{Muonic hydrogen atom beams from thin films}

The basic processes involved in creating a beam of muonic hydrogen
atoms can be categorized into four step:\cite{marku96} {\em atomic
formation}, {\em acceleration}, {\em extraction}, and {\em
moderation}. Let us take as an example the production of muonic
tritium atom ($\mu t$) beam.\cite{fujiw97} When a muon is stopped
in a thin solid hydrogen target consisting of protium ($^1$H$_2$)
doped with a small amount ($c_t\sim 0.1$\%) of tritium, muonic
protium atom ($\mu p$) is mostly formed ({\em atomic formation}).
The muon quickly transfers from proton to triton\cite{mulha96} to
form $\mu t$, the latter more tightly bound due to the reduced
mass difference. In the reaction, the $\mu t$ gains about 45 eV of
recoil kinetic energy, and is thus {\em accelerated}. The $\mu t$
then slow down from the collisions with the rest of the target
nuclei (mostly protons), until it reaches about 10 eV. At these
energies, $\mu t + p$ elastic scattering cross section drops
dramatically due to a phenomenon known as the Ramsauer-Townsend
effect,\footnote{Elastic scattering cross section goes to zero,
when the $s$-wave phase shift goes to $\pi$ and the contributions
from higher partial waves are negligible.} making the rest of the
target nearly transparent. The $\mu t$ atom is thus {\em
extracted} from the layer into vacuum. The energy of the emitted
$\mu t$ can be controlled to some extent by placing an additional
layer, for example of deuterium, on top of the emission layer
({\em moderation}). Creation of muonic deuterium ($\mu d$) is
possible in a similar manner with a deuterium-doped protium
target: in fact, the emission of muonic hydrogen atoms was first
discovered in this system.\cite{forst90}

Unexpected emission of $\mu p$ from a pure $^1$H$_2$ target was
observed recently,\cite{wozni99} which indicates the existence of
unconventional acceleration and extraction mechanism. Presumably,
$\mu p$ is accelerated during the atomic cascade\cite{marku99} to
the ground state, while $\mu p$ extraction from the layer may be
enhanced due to the condensed matter effects\cite{adamc99,knowl97}
in $\mu p + p$ scattering. The energy of emitted $\mu p$, though
it has not been directly measured, is expected to be of order
$\sim$meV given by the Bragg cut-off limit, and its hyperfine
state to be in $F=0$ due to the high spin exchange rate in the
target.

These observations of $\mu p$, $\mu d$, and $\mu t$ emission are
not only useful for new type of experiments, but also interesting
in its own right in testing the quantum few body calculations.
Because of the large mass of the muon, comparable to that of
nuclei, reactions of muonic atoms is non-adiabatic in nature, and
the corrections due to relativistic and QED effects may be
non-negligible in some cases. Measurements reported up to now by
the TRIUMF group include, charge exchange
processes\cite{mulha96,jacot96} $\mu p + d \rightarrow p + \mu d$,
$\mu p + t \rightarrow p + \mu t$; elastic scattering
processes\cite{mulha99,fujiw99} $\mu d + p \rightarrow \mu d + p$,
$\mu t+ p \rightarrow \mu t + p$; spin exchange
process\cite{knowl97} $\mu d^{F=\frac{3}{2}} + d \rightarrow \mu
d^{F=\frac{1}{2}}+d$; molecular formation\cite{mulha96,knowl97}
$p\mu p$, $d\mu d$; and the astrophysical $S$ factor from $p\mu d$
fusion.\cite{olin} Many of these processes play important roles in
$\mu$CF as well as in other fundamental muon physics experiments.
In the following sections, we focus on two crucial processes,
$d\mu t$ formation and $\mu$-$\alpha$ sticking.

\section{Resonant molecular formation in $\mu$CF}

Muon catalyzed fusion in the deuterium-tritium system currently
has two major bottle-necks in term of achieving high efficiency.
One is the rate at which a muon can go through the catalysis cycle
(cycling rate), and another is a poisoning process called
$\mu$--$\alpha$ sticking in which, with a probability $\omega _s
\sim 0.005$, the muon gets captured to atomic bound states of the
fusion product $^4$He after the fusion reaction, and hence lost
from the cycle (see Section 4). The former is limited mainly by
the rate of formation of muonic molecule $d\mu t$. In fact, a
straightforward mechanism for molecular formation via Auger
process is much too slow, yielding the fusion efficiency of the
order of only one fusion per muon. A resonant mechanism, however,
can give much higher rates when the certain condition is
satisfied. In the resonant formation, $\mu t + D_2 \rightarrow
[(d\mu t)dee]_{\nu K}$, the reaction product is a hydrogen-like
molecular complex $[(d\mu t)dee]$ in which $(d\mu t)^+$ plays a
role of one of the nuclei. The process is resonant, because the
energy released upon formation of $d\mu t$ plus the $\mu t$
kinetic energy has to coincide the ro-vibrational ($\nu K$)
excitation spectrum of $[(d\mu t)dee]$ in the final state. This is
only possible due to the existence of a state in $d\mu t$ which is
bound very loosely (in the muonic scale) with the binding energy
smaller than the dissociation energy of deuterium molecules.

\begin{figure}[t]
 \begin{center}
 \epsfig{file=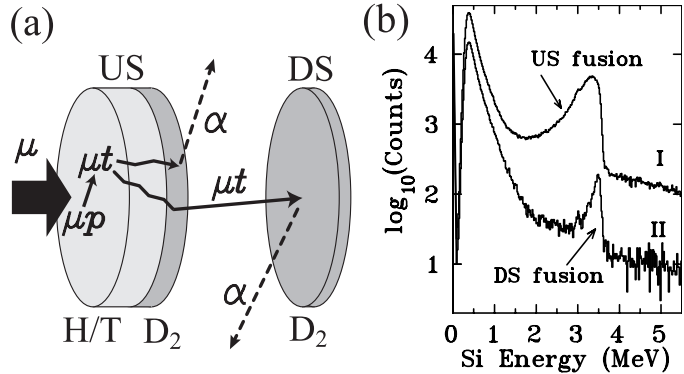,width=0.75\textwidth}
  \end{center}
\caption{(a) Schematic of the thin film target for the $d\mu t$
formation measurements consisting of the emission, the moderation
and the reaction layers, which are prepared by rapidly freezing
hydrogen isotopes on the gold foils (not shown) held in vacuum at
3.5 K.\protect\cite{knowl96} The layer thickness (3.43
mg$\cdot$cm$^{-2}$, 96 $\mu$g$\cdot$cm$^{-2}$, and 21
$\mu$g$\cdot$cm$^{-2}$, respectively) were measured off-line via
$\alpha$ particle energy loss.\protect\cite{fujiw97b} (b) Measured
Si energy spectra with prompt (I: $t>0.02$ $\mu$s) and delayed
(II: $t>1.5$ $\mu$s) time cuts. Fusion in DS reaction layer is
separated from that in US D$_2$ due to the $\mu t$ TOF across the
vacuum.}
 \label{fig:siE}
\end{figure}

Theoretical calculations\cite{faifm91,petrov} predict strong
enhancement of the formation rate $\lambda _{d\mu t}$ at $\mu t$
energy of order 1 eV, but direct experimental information on such
epithermal resonance is scarce. In conventional $\mu$CF
experiments, determination of $\lambda _{d\mu t}$ is rather
indirect and model dependent due to the complexity of muonic
reaction cycles in the target. As well, the resonant energies
$\sim$1 eV is difficult to access with the target thermal energy,
since it would require a target of several thousand degrees, a
formidable task when working with tritium. The use of $\mu t$ beam
provides some unique advantages in this case: (1) formation
process can be isolated, to large extent, from the rest of the
cycle, (2) epithermal energies are directly accessible due to the
available beam energy, and (3) $\mu t$ time of flight across the
drift distance provide information of the resonance energies.

Figure~\ref{fig:siE} (a) illustrates the principle of our
method.\cite{marsh92} A beam of $5\times 10^3$ $\mu ^-$/s of
momentum 27 MeV/$c$ from the TRIUMF M20B channel was degraded in a
51 $\mu$m gold target support, and stopped in the upstream (US)
emission layer. The $\mu t$ beam, obtained as described above,
were slowed via elastic scattering in a D$_2$ moderation layer
from some 10 eV to near 1 eV to better match the resonance
energies. The $\mu t$, after a few $\mu$s time of flight (TOF), is
collided with a reaction layer in downstream (DS), separated by
the drift distance of 17.9 mm in vacuum. Formation of $d\mu t$
molecules is detected by observing 3.5 MeV $\alpha$ particles
produced in the fusion reaction, $d + t \rightarrow \alpha + n$,
which follows the formation. Si detectors placed in the vacuum
enables the measurement $\alpha$ with high energy resolution. The
background can be determined accurately by ``turning off'' the DS
fusion reactions using the target without the DS layer. This
ability to control a  specific process, without affecting the
rest, is an advantage of the thin film method. (In conventional
targets, changing the target conditions would affect many
processes simultaneously).

Because the time between muon stop and the fusion $\alpha$
detection is dominated by the $\mu t$ TOF, it provides a measure
on molecular formation energy, as long as the energy loss of $\mu
t$, due to elastic scattering before the formation in the DS
D$_2$, is small. A thin DS layer ($\sim$1 $\mu$m)
was necessary to minimize this energy loss so as not to obscure
the time-energy correlation.

 \begin{figure}[t]
 \begin{center}
 \epsfig{file=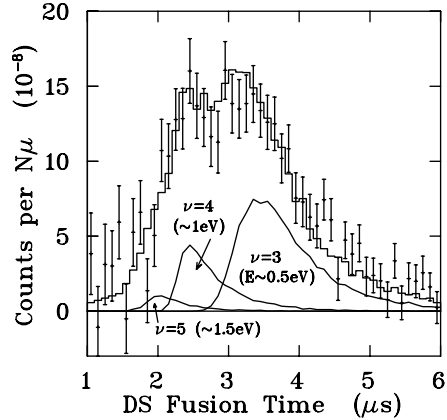,width=0.48\textwidth}
 \end{center}
 \caption{Time-of-flight fusion spectrum (error bars) and
 simulation spectrum (histogram), normalized the number of
 incident muons $N_\mu$. Also plotted are simulated
 contributions from different resonance peaks given by
 the time--energy correlated events.}
 \label{fig:mc}
 \end{figure}

The detail of our analysis can be found in Refs.\cite{phd,prl} and
here we give the resulting DS fusion time spectrum and its
comparison with Monte Carlo (MC) simulations in Fig.~\ref{fig:mc},
which clearly establishes the resonance structure. From the
time-of-flight analysis of $2036\pm 116$ DS fusion events, a
formation rate consistent with $0.73\pm (0.16)_{meas} \pm
(0.09)_{model}$ times the theoretical prediction of Faifman {\it
et al.}\cite{faifm91} was obtained (the first error refers to the
measurement uncertainty including the statistics and the second is
that in MC modeling). The resonance energies were determined from
the fit to be $0.940\pm (0.036)_{meas} \pm (0.080)_{model}$ times
the theory.\cite{faifm91} Thus, for the first time, the existence
of epithermal resonances in $d\mu t$ molecular formation was
directly confirmed, and their energies measured. For the largest
peak at the resonance energy of $0.423\pm 0.037$ eV, our results
correspond to the peak rate of $(7.1 \pm1.8)\times 10^9$ s$^{-1}$.
This is more than an order of magnitude larger than the rates at
lower energies, experimentally demonstrating the prospect for high
cycling $\mu$CF in a high temperature target of several thousand
degrees. If one assumes the energy levels of the [$(d\mu t)dee$]
molecule, which have a similar structure to those of ordinary
hydrogen, our results imply sensitivity to the binding of energy
of the loosely bound $(d\mu t)_{11}$ state with an accuracy
comparable to the vacuum polarization and other QED corrections,
opening up a new possibility of precision spectroscopy in a
quantum few body system.

The data have been subsequently collected\cite{porce99} for $d\mu
t$ formation in the $\mu t + HD$ collision, for which for stronger
resonances at lower energies are predicted. The result will be
reported in a future publication.

\section{Direct measurement of $\mu$--$\alpha$ sticking}

The process of $\mu$-$\alpha$ sticking give a stringent limit,
independent of the cycling rate, on the number of fusions
catalyzed by one muon. As such, great efforts have been made for
nearly two decades to understand this process, but discrepancies
persist between and theory and experiment (including latest
PSI\cite{petit93} and RIKEN-RAL\cite{riken} results), the latter
being systematically lower than the former.

 \begin{figure}[t]
 \begin{center}
 \epsfig{file=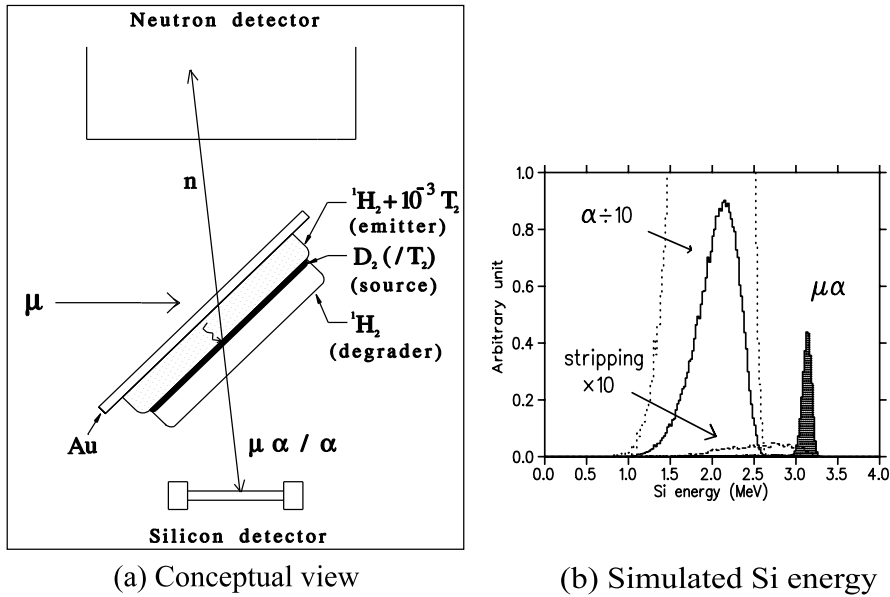,width=\textwidth}
 \end{center}
 \caption{Proposed direct measurement of $\mu$-$\alpha$
 sticking.\protect\cite{fujiw96,eec}
 (a) The $\mu t$ created in the {\it emitter} is stopped
 in the {\it source}, where fusion takes place producing $\alpha ^{++}$
 or ($\mu \alpha$)$^+$. The {\it degrader} separates the two species
 by their stopping powers (recall that $dE/dx \propto
 Z^2$). Collinear coincidence of the neutron with the charged events
 suppresses the background. (b) Simulated Si
 spectrum for the coincidence events, showing a clear separation
 between $\alpha$ and $\mu \alpha$ peaks.}
 \label{fig:stick}
 \end{figure}

Calculations of sticking are challenging, due to the interplay of
the Coulomb and strong interactions in a non-adiabatic few-body
system, yet recent predictions, including the effects of nuclear
structure and the deviations from the standard sudden
approximation, They cannot, however, be readily compared to
experiment because most measurements are primarily sensitive to
{\it final sticking} ($\omega _s^{fin}$), which is a combination
of {\it initial sticking} ($\omega _s^{0}$), the intrinsic
branching ratio for $d\mu t \rightarrow \mu \alpha +n$, and {\it
stripping} ($R$), collisional reactivation of the muon from $\mu
\alpha$ in the target medium ({\it i.e.}, $\omega _s^{fin} \equiv
\omega _s^{0} (1-R)$).

Confusing history\cite{review} of experimental sticking results
has been in part attribu\-ted\cite{petit93} to the difficulties
associated with the conventional neutron method, which include a
high model dependence and the need for the absolute neutron
calibration. A recent RIKEN-RAL experiment\cite{riken} has
directly observed, with impressive statistics, X-rays from $\mu
\alpha$ deexcitaion\footnote{Upon sticking a fraction of the muon
gets captured into excited atomic states of $\mu \alpha$ and can
emit X-rays if it deexcites to the ground state.} taking advantage
of the new intense pulsed muon beam which enabled the large signal
enhancement, but unfortunately the determination of the sticking
probability has to rely on the models of $\mu \alpha$ cascade and
stripping, which are yet largely untested.

We propose a new direct experiment of sticking using multi-layer a
thin film target. The method is illustrated in
Fig~\ref{fig:stick}, and the details are given in
Refs.\cite{fujiw96,eec} The determination of sticking from the
ratio $\mu \alpha / (\mu \alpha + \alpha)$ is simple and model
independent. Since the stripping in the degrader is small, the
measurement is sensitive to the initial sticking, while the
stripping itself process can be systematically studied by varying
the degrader thickness. Thus experimental separation of initial
sticking and stripping will become possible for the first time.
The use of very thin ($<$ few $\mu$m) source layer is essential
for minimizing the peak broadening in order to achieve good energy
separation of the $\mu \alpha$ and $\alpha$. The expected
precision of $\sim$5\% in initial sticking is limited by the
statistics.

\section{Summary}
The thin film methods have yielded unique results in the studies
of muonic atoms and muon catalyzed fusion, and offer opportunities
for more new experiments. These measurements, because of their
thin targets, could further benefit from the advent of high
intensity, high quality muon sources.

\section*{Acknowledgements}
This work is supported in part by Canada's NSERC. MCF acknowledges
the support of the Government of Canada, Green College, Nortel,
and JSPS.

\end{document}